\documentclass[aps,rsi,preprint,groupedaddress]{revtex4-1}
\usepackage{dcolumn}
\usepackage{amsmath}
\usepackage{color}
\usepackage{subfigure}
\usepackage{graphicx}
\begin{document}
\title{Low magnetic Johnson noise electric field plates for precision measurement}

 \author{I. M. Rabey} \author{J. A. Devlin} \author{E. A. Hinds} \author{B. E. Sauer}

\affiliation{Centre for Cold Matter, Blackett Laboratory, Imperial College London, Prince Consort Road, London SW7 2AZ UK}

\email[]{ben.sauer@imperial.ac.uk}

\date{\today}

\begin{abstract}
We describe a parallel pair of high voltage electric field plates designed and constructed to minimise magnetic Johnson noise.  They are formed by laminating glass substrates with commercially available polyimide (Kapton) tape, covered with a thin gold film. Tested in vacuum, the outgassing rate is less than $5\times10^{-5}$ mbar.l/s. The plates have been operated at electric fields up to 8.3 kV/cm, when the leakage current is at most a few hundred pA. The design is discussed in the context of a molecular spin precession experiment to measure the permanent electric dipole moment of the electron.
\end{abstract}

\pacs{1.23a}

\maketitle


\newcommand{\Eeff}{E_\text{\footnotesize eff}}
\newcommand{\msub}[1]{\text{\footnotesize #1}}

\section*{Introduction}

Spin precession measurements on atoms, molecules or solid state samples are susceptible to noise from magnetic field fluctuations. For measurements performed close to the surface of conductors, the magnetic Johnson noise caused by thermal fluctuation of electrons in the conductor \cite{varpula84} can contribute significantly to the uncertainty of the measurement. This effect is a known challenge in atom chip experiments \cite{hinds03, haroche09} and precision magnetometry \cite{toiro96}. Magnetic Johnson noise is also becoming a challenge for experiments that measure the electron's electric dipole moment (EDM) \cite{lamoreaux99, munger05, amini07}. Measurements of the electron EDM are typically carried out in neutral atoms or molecules \cite{commins05, hudson11, acme14} using some form of electron spin precession. Limitations associated with magnetic noise in EDM experiments can be avoided by choosing neutral system  with a small magnetic moment and/or a large sensitivity to the size of EDM \cite{cornell11}. Even using a favorable system, magnetic Johnson noise must be minimised through careful choice of the materials placed close to the atoms or molecules. This paper describes a simple, low cost set of electric field plates which will be used for an improved measurement of the electron EDM using the YbF molecule \cite{devlin15}.

The YbF experiment measures the phase difference between Zeeman sublevels of the molecule in an applied electric field $E$ (which defines the $z$-axis) and an applied magnetic field  $B$. The phase shift $\phi$ of a level has an electric part $\phi_E = d_e \Eeff\tau/\hbar$, where $d_e$ is the electron's EDM and $\Eeff$ is the effective electric field interacting with the EDM, and $\tau$ is the duration of the interaction.  There is also a magnetic part $\phi_B = \int_0^\tau dt\, \mu B_z(t)/\hbar$, where $\mu$ is the magnetic moment (in our case the Bohr magneton) and  $B_z$ is the magnetic field component parallel to $E$. Details of the experiment are given in \cite{kara12}. The uncertainty in $d_e$ should ideally be limited by the shot noise associated with the number of YbF molecules detected, but $\phi$ also has noise due to the fluctuations of $B_z$. In order to benefit from current improvements that lengthen $\tau$ and increase the number of detected molecules, it is now becoming important to reduce the magnetic noise, which is dominated by the Johnson noise due to current fluctuations in the electric field plates \cite{lamoreaux99,munger05}.

\section*{Magnetic Johnson noise in the YbF EDM experiment}

Let the time-dependent noise in $B_z$ be $B_{n,z}(t)$. We relate this to the corresponding spectral density of the noise power, $B^2_{n,z}(f)$, by the Wiener-Khintchine relation
\begin{equation}
\left\langle B_{n,z}(t_0)B_{n,z}(t_0+t)\right\rangle=\int_{0}^{\infty} df\,B^2_{n,z}(f)\cos(2\pi f t)\,.
\label{Eq:WKTheorem}
\end{equation}
The rms phase uncertainty after interaction time $\tau$ is 
\begin{equation}
\sigma_{\phi_{\rm B}}(\tau)= \frac{\mu}{\hbar} \left\langle\left(\int_{t_0}^{t_0+\tau} B_{n,z}(t)\,\textrm{d}t\right)^2\right\rangle^{1/2}. 
\label{Eq:phaseErrEq}
\end{equation}
As noted in \cite{munger05}, we can use Eq.~(\ref{Eq:WKTheorem}) to rewrite Eq. (\ref{Eq:phaseErrEq}) in the form
\begin{equation}
\label{eq:mungerD}
\sigma_{\phi_{\rm B}}(\tau)= \frac{\mu \tau}{\hbar}\left(\int_{0}^{\infty}B^{2}_{n,z}(f)\frac{\sin^{2}(\pi f\tau)}{(\pi f \tau)^{2}}~df\right)^{1/2}\,.
\end{equation}
Eqation (38) of Varpula and Poutanen \cite{varpula84}  provides a formula for $B^{2}_{n,z}(f)$ at distance $z$ from an infinite slab of conducting material at temperature $T$, having thickness $d$ and resistivity $\rho$.  In reality the molecule passes between two electric field plates, each contributing to the magnetic field noise, so we multiply Eq.(38) of \cite{varpula84} by a factor of $\sqrt{2}$:
\begin{equation}
\label{eq:bnz}
B_{n,z}(f)=2\mu\sqrt{\frac{k_{B}T}{\pi\rho}}\left(\int_{0}^{\infty}R(\xi,t)e^{-2\xi z}\xi~d\xi\right)^{1/2} \, ,
\end{equation}
 where $R(\xi,t)$ is a complicated expression given in Eq.(39) of \cite{varpula84}. The $\xi $ variable encodes the frequency dependence of the magnetic field noise via Fourier integrals. 
 
\begin{figure}[t]
\centering
\includegraphics[scale=1]{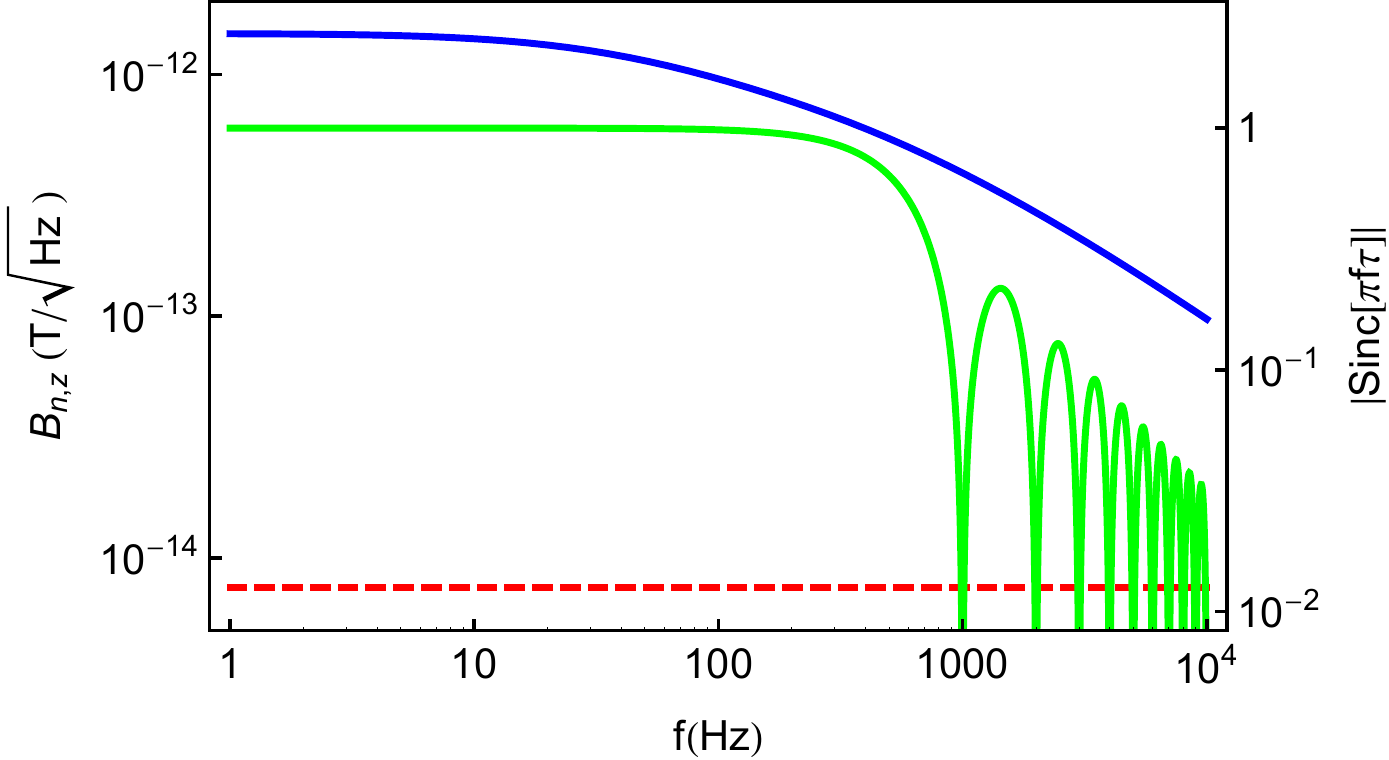}
\caption{\label{fig:bnzplot}Quantities to be integrated in Eq.(\ref{eq:mungerD}). Blue: magnetic field noise spectrum $B_{n,z}(f)$ at a temperature of $300$~K, calculated using eq. \ref{eq:bnz} between aluminium plates (${\rho=2.7\times 10^{-8}~\Omega\,}$m) of thickness $d=1.27$~cm and separation $2z=12$~mm. Green: the function $|\text{sinc}(\pi f \tau)|$, taking an interaction time $\tau$ of 1~ms. Red (Dashed): $B_{n,z}(f)$ between gold film plates ($\rho=2.3\times 10^{-8}~\Omega\,$m) of $90$~nm thickness.}
\end{figure}

The last generation of the YbF eEDM experiment \cite{hudson11, kara12} used aluminium electric field plates (${\rho=2.7\times 10^{-8}\,\Omega\,}$m)  of thickness $d=12.7$~mm. The gap between the plates was 12 mm, so the distance from the molecules to the surface was $z=6$ mm. At that distance, the magnetic noise spectrum given by Eq.(\ref{eq:bnz}) is plotted as the top (blue) line in Fig.~\ref{fig:bnzplot}. The noise tends to a constant at low frequency, given by \cite{varpula84}
\begin{equation}
\label{eq:bnzf0}
B_{n,z}(0)=\mu_0\sqrt{\frac{k_{B}T}{4\pi\rho}\frac{d}{z(z+d)}}\,.
\end{equation}
The integrand in Eq.(\ref{eq:mungerD}) also involves the function ${\rm sinc}(2\pi f \tau)$, shown (green, middle) in Fig.~\ref{fig:bnzplot}. Here ${\tau=1}$~ms as that was the interaction time in the experiment. After squaring and multiplying these two  functions, numerical integration of  Eq.(\ref{eq:mungerD}) gives the phase noise as $\sigma_{\phi_B}=1.5$~mrad. This causes an uncertainty in the electron EDM of $\sigma_{B}\hbar/(\tau\Eeff)=8 \times 10^{-26}$ e.cm for a single measurement over coherence time $\tau~=~1$~ms in an electric field of 8.3 kV/cm. Averaging over a data run of $4\times 10^7$ measurements the corresponding Johnson noise in $d_e$ is $1.2 \times 10^{-29}$ e.cm. While this was negligible in the previous version of the experiment, that is no longer the case \cite{devlin15,Hinds13}, so we have designed new plates with a much lower Johnson noise. 

Equation (\ref{eq:bnzf0}) suggests four ways to reduce the magnetic noise: reduce $T$, increase $z$, increase $\rho$, and reduce $d$. The temperature does not offer a significant improvement unless the plates could be cooled far below liquid nitrogen temperature, and this does not seem practical in a molecular beam apparatus. Larger plate spacing is also not a very useful option for the YbF as it would be challenging to maintain the required high electric field.  

With a view to increasing $\rho$, we have built field plates from commercial ITO coated glass (Pilkington K glass), which has good vacuum properties and is transparent to visible light. We find that the coating is prone to discharge, particularly at the edges. This problem was also reported by \cite{solmeyer13}, who tried  to mitigate it by custom coating ITO over rounded edges. Also, the surface is very easily damaged by discharges. It has also been proposed that one might use soda glass, heated to a temperature of 150$^{\circ}$C to achieve enough conductivity \cite{amini07}, but the addition of plate heaters is a complication we prefer to avoid in the YbF experiment. 

Our approach has therefore been to reduce $d$, which we have done by making the electrodes from gold-coated polyimide (Kapton) tape \cite{tape}, on which the gold layer is only $90$~nm thick. This is roughly $10^{-4}$ of the previous plate thickness, while the $2.3\times 10^{-8}\,\Omega\,$m resistivity of gold is only slightly less than that of aluminium. Equation~(\ref{eq:bnzf0}) suggest that this will reduce the magnetic field noise by  approximately 100, and numerical integration of Eq.(\ref{eq:bnz}) shows this to be the case, as plotted by the red dashed line in Fig.\ref{fig:bnzplot}. The corresponding phase uncertainty, given by Eq.(\ref{eq:mungerD}), is  $15\,\mu$rad for a 1~ms measurement in the YbF experiment, which converts to an uncertainty in the electron EDM of $1.1\times 10^{-31}$~e.cm over a data run of $4\times 10^7$ measurements. This is sufficiently small that the new YbF experiment will not be limited by magnetic noise from the new plates. We note that the magnetic noise from the glass substrates is negligible \cite{henkel99}.

\section*{Construction and testing of the plates}

\begin{figure}[b]
\centering
\includegraphics[scale=1]{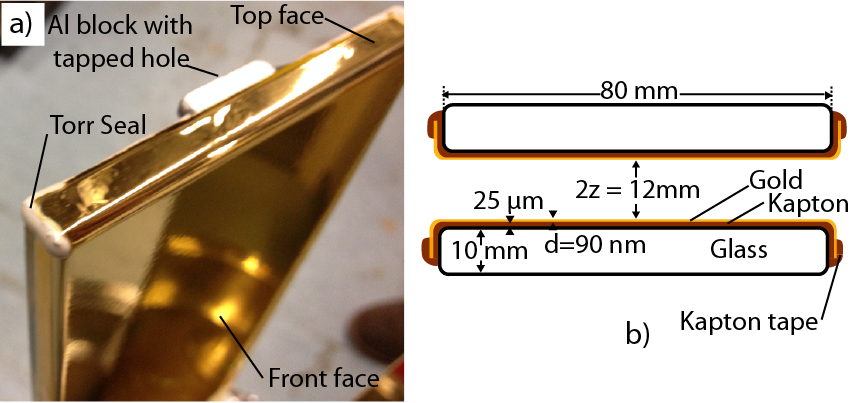}
\caption{\label{fig:plates}Photograph and cross section of the film-on-glass electric field plates.}
\end{figure}

These new, thin, field plates are built on slabs of float glass, with dimensions $835\times 80 \times10$~mm, chamfered on the edges at 45$^\circ$ by a few mm, and rounded by hand using silicon carbide abrasive paper. These are cleaned using detergent and water, then rinsed in distilled water followed  by isopropanol. A cold-roll laminator applies a continuous length of the self-adhesive, gold-coated polyimide tape to the front face of each plate, as shown in Fig. \ref{fig:plates}(a). Excess tape is then folded over onto all four sides of the glass, as illustrated in cross-section in Fig. \ref{fig:plates}(b). When viewed in daylight the surface shows a ripple with a period of a few millimetres, but on measuring the height variations on a flat bed we found less than $\pm\,2.5\mu$m peak-to-peak over a $50$~mm-square region and no perceptible ripple.

On the top side, viewed in Fig.\ref{fig:plates}(a), the tape is folded over to the back, where it is cut to form a small tab of $5 \times 15$ mm, with its long axis parallel to the top edge of the substrate. An aluminium block, shown in Fig.\ref{fig:plates}(a), is glued over this tab using conducting epoxy, and is tapped to accept the threaded 2-mm end of a vacuum-compatible high-voltage cable. The exposed edges of the gold-coated tape discharge at high voltage, but this is suppressed when we cover them with uncoated self-adhesive Kapton tape. The four short corners are covered with a small quantity of \emph{Torr Seal} vacuum epoxy to avoid any awkward edges.

Inside the vacuum chamber, the plates are mounted vertically along the beam line. Each plate is held by two horizontal alumina rods, glued with  structural epoxy \cite{scotch} to the back face at the top and the bottom. These plug into two aluminium support plates, where precisely machined holes set the plate separation. The length of the leakage path from the gold conductor to the grounded aluminium support structure is 21 mm. 

Once assembled and placed in the vacuum chamber, the entire apparatus is baked at 80$^{\circ}$C for three days to accelerate the outgassing. After cooling to room temperature, the pressure in the vacuum chamber falls to $4\times 10^{-7}$ mbar, which is also the base pressure of the chamber without the plates. We conclude that the outgassing rate from the tape with its acrylic adhesive is below $5\times10^{-5}$ mbar.l/s. 

Next, the plates are conditioned to support a large potential difference by applying a voltage and switching its polarity with about a one minute period. We monitor the leakage current using nA-sensitive leakage monitors \cite{sauer08} and raise the voltage in 100V steps until current spikes of $10-100$~nA start to appear. Then the voltage is held constant (while continuing to switch the polarity) until the spikes disappear. This takes between a few minutes and a few hours, after which the voltage can be further increased. If the leakage current becomes continuous at greater than 10nA the voltage is reduced for a time. If large leakage currents ($> 1\mu$A) are observed the voltage is reduced to zero and conditioning starts afresh. Unlike the ITO coating, the gold suffered no perceptible damage from these discharges, perhaps because the gold has much higher conductivity, both electrical and thermal. 

With this procedure, we reach a voltage of $\pm 5$kV after several days of conditioning. The leakage current from the plates and high voltage cabling is $0.4\pm0.1$ nA, which is dominated by leakage in the external 10m-long coaxial connecting cables and high voltage switching equipment. The leakage current from the plates and high voltage connections under vacuum is less than 200 pA. 

\begin{figure}[h]
\centering
\includegraphics[scale=1]{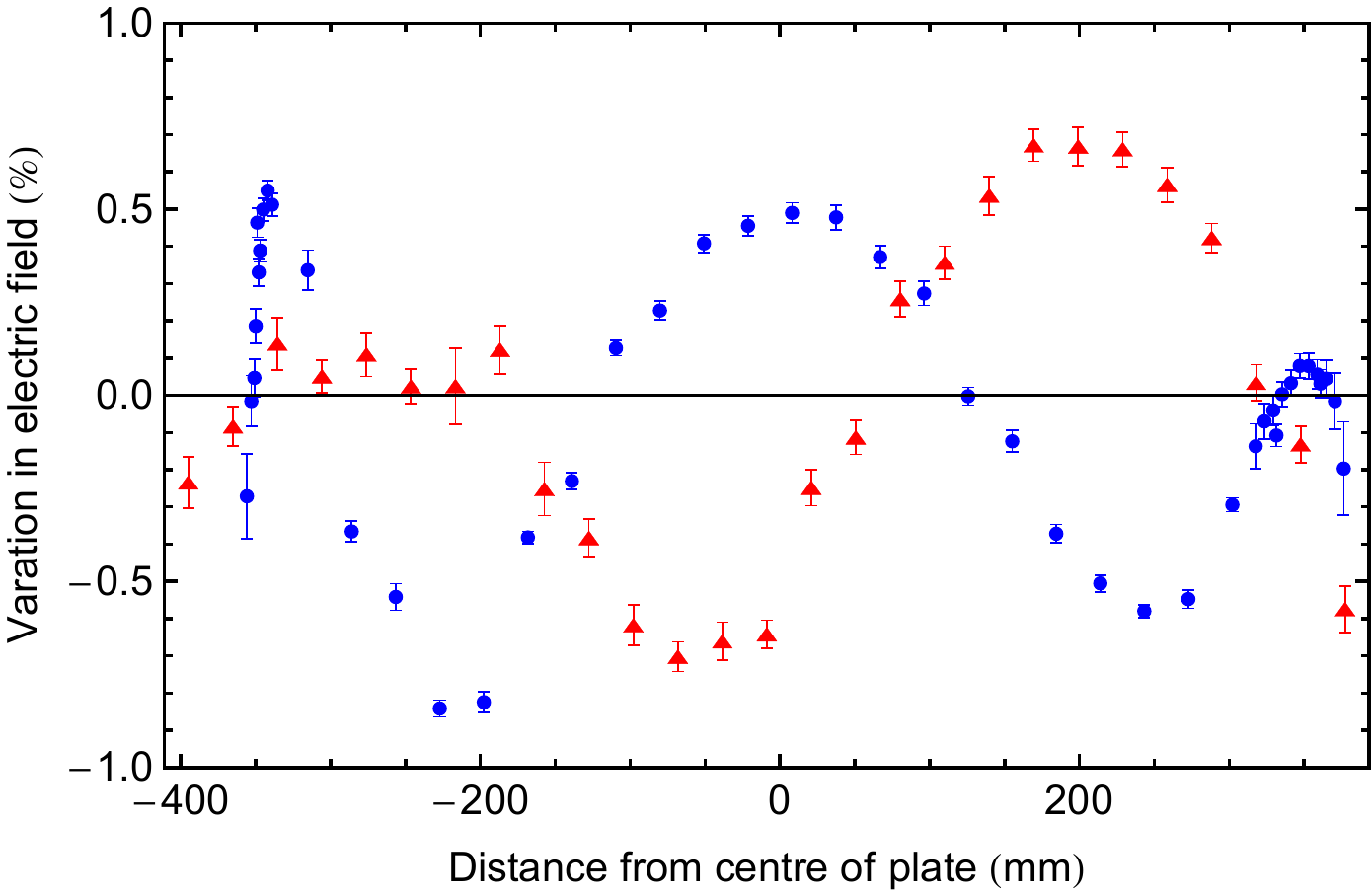}
\caption{\label{fig:separation}Fractional variation of the electric field as a function of distance along the beam line, as measured by the Stark shift of the hyperfine structure in YbF. Blue circles: aluminium plates. Red triangles:  tape-on-glass plates.}
\end{figure}

It is important in the EDM experiment that the molecules experience a homogeneous electric field between the plates \cite{Tarbutt09, kara12}. We have mapped the field between the new plates at 4\,kV/cm, using the molecules themselves to sense the field strength through the Stark effect of the hyperfine structure \cite{Condylis05}. The triangular data points (red online) in Fig.~\ref{fig:separation} show the fractional variation of the field as a function of position along the length of the plates. This  is within the range $\pm 0.8\%$, which provides a sufficiently uniform Stark shift for the EDM experiment. For comparison, we also show the fractional field variation measured between the old plates, which is very similar in magnitude. Rotation of the electric field direction can also be undesirable when it induces a Berry phase \cite{Tarbutt09, kara12}, but the variation we measure here occurs over a long distances and that implies a small rotation angle that is insufficient to produce a significant systematic error in the EDM experiment.

It would also have been been possible to apply the gold layer by vacuum deposition, but the tape is considerably cheaper and allows field plates of any length to be produced. It is also straightforward to replace the tape in-house if it were to become damaged by electrical discharge or surface contamination.

\section*{Acknowledgements}
We acknowledge useful discussions with M. Tarbutt. The research leading to these results has received funding from EPSRC, STFC, the Royal Society and from the European Research Council under the European Union’s Seventh Framework Programme (FP7/2007-2013) / ERC grant agreement 320789.”


\end{document}